\documentclass[twoside]{article}
\usepackage{amsfonts,amssymb,amsbsy,textcomp,marvosym,amsmath,caption,threeparttable,amsthm,subcaption,float,lastpage,lscape,siunitx}
\usepackage{eurosym,mathrsfs,fancyhdr,picins,CJK,multicol,graphics,indentfirst,color,bm,upgreek,booktabs,graphicx,multirow,warpcol,natbib}
\usepackage[bookmarksnumbered=true,bookmarksopen=true,colorlinks=true,pdfborder=001,urlcolor=blue,linkcolor=blue,anchorcolor=blue,citecolor=blue]{hyperref}

\usepackage[capitalize]{cleveref}

\newcommand{\eg}{e.\,g.}
\newcommand{\ie}{i.\,e.}

\newcommand{\db}{\si{\deci \bel}}

\usepackage{soul}
\usepackage{epstopdf}
\usepackage{trackchanges}

\usepackage{cleveref}
\usepackage{todonotes}
\usepackage[acronym, shortcuts, nohypertypes={acronym}]{glossaries}
\usepackage[inline]{enumitem}
\newacronym{AE}{AE}{audio enhancement}
\newacronym{AI}{AI}{artificial intelligence}
\newacronym{ASR}{ASR}{automatic speech recognition}
\newacronym{DCASE}{DCASE}{Detection and Classification of Acoustic Scenes and Events}
\newacronym{SCR}{SCR}{speech command recognition}
\newacronym{ASC}{ASC}{acoustic scene classification}
\newacronym{SED}{SED}{sound event detection}
\newacronym{CA}{CA}{computer audition}
\newacronym{CAT}{CAT}{computer audition task}
\newacronym{UAR}{UAR}{unweighted average recall}
\newacronym{CNN}{CNN}{convolutional neural network}
\newacronym{DL}{DL}{deep learning}
\newacronym{DNN}{DNN}{deep neural network}
\newacronym{SE}{SE}{speech enhancement}
\newacronym{SER}{SER}{speech emotion recognition}
\newacronym{SNR}{SNR}{signal-to-noise ratio}
\newacronym{WER}{WER}{word error rate}

\renewcommand{\thefootnote}{\fnsymbol{footnote}}
\def\footnoterule{\kern 1mm \hrule width 10cm \kern 2mm}

\allowdisplaybreaks
\catcode`@=11
\def\title#1{\vspace{3mm}\begin{flushleft}\vglue-.1cm\Large\bf\boldmath\protect\baselineskip=18pt plus.2pt minus.1pt #1
\end{flushleft}\vspace{1mm} }
\def\author#1{\begin{flushleft}\normalsize #1\end{flushleft}\vspace*{-4pt} \vspace{3mm}}
\def\address#1#2{\begin{flushleft}\vglue-.35cm${}^{#1}$\small\it #2\vglue-.35cm\end{flushleft}\vspace{-2mm}\par}

\catcode`@=11
\def\section{\@startsection{section}{1}{\z@}%
 {-3ex \@plus -.3ex \@minus -.2ex}%
 {2.2ex \@plus.2ex}%
{\normalfont\normalsize\protect\baselineskip=14.5pt plus.2pt minus.2pt\bfseries}}
\def\subsection{\@startsection{subsection}{2}{\z@}%
 {-3ex\@plus -.2ex \@minus -.2ex}%
 {2ex \@plus.2ex}%
{\normalfont\normalsize\protect\baselineskip=12.5pt plus.2pt minus.2pt\bfseries}}
\def\subsubsection{\@startsection{subsubsection}{3}{\z@}%
 {-2.2ex\@plus -.21ex \@minus -.2ex}%
 {1.4ex \@plus.2ex}
{\normalfont\normalsize\protect\baselineskip=12pt plus.2pt minus.2pt\sl}}

\begin{document}
\begin{CJK*}{GBK}{song}
\thispagestyle{empty}
\vspace*{-13mm}
\noindent {\small Manuel Milling, Shuo Liu, Andreas Triantafyllopoulos {\it et al.} Journal of computer science and technology: Instruction for authors.
JOURNAL OF COMPUTER SCIENCE AND TECHNOLOGY \ 33(1): \thepage--\pageref{last-page}
\ January 2018. DOI 10.1007/s11390-015-0000-0}
\vspace*{2mm}

\title{Audio Enhancement for Computer Audition \\ -- An Iterative Training Paradigm Using Sample Importance}

\author{Manuel \textcolor{blue}{Milling}$^{1,2,3,*}$, Shuo Liu$^{1}$ \begin{CJK*}{UTF8}{gbsn}(刘硕)\end{CJK*},
Andreas Triantafyllopoulos$^{1,2,3}$, Ilhan Aslan$^{4}$,
and Bj\"{o}rn W.\ Schuller$^{1,2,3,5,6}$}

\address{1}{EIHW -- Chair of Embedded Intelligence for Health Care \& Wellbeing, University of Augsburg, Germany}
\address{2}{CHI -- Chair of Health Informatics, MRI, Technical University of Munich, Germany}
\address{3}{MCML -- Munich Center for Machine Learning, Germany}
\address{4}{Huawei Technologies, Munich, Germany}
\address{5}{MDSI -- Munich Data Science Institute, Germany}
\address{6}{GLAM -- the Group on Language, Audio,
\& Music, Imperial College London, UK}

\vspace{2mm}

\noindent E-mail: manuel.milling@tum.de; shuo.liu@informatik.uni-augsburg.de; andreas.triantafyllopoulos@tum.de; ilhan.aslan@huawei.com; schuller@tum.de\\[-1mm]

\noindent Received October 26, 2022; accepted March 08, 2024.\\[1mm]

\let\thefootnote\relax\footnotetext{{}\\[-4mm]\indent\ Regular Paper\\[.5mm]
\indent\ by the National Natural Science Foundation of China under Grant Nos.~******** and ********, the National High Technology Research and Development 863 Program of China under Grant No.~********, the National Basic Research 973 Program of China under Grant No.~********, and the Natural Science Foundation of Shandong Province of China under Grant No.~*******. \\[.5mm]
\indent\ $^*$Corresponding Author
\\[1.2mm]\indent\ $^{\footnotesize\textcircled{\tiny1}}$\href{https://jcst.ict.ac.cn/Guidelines_for_Authors}{https://jcst.ict.ac.cn/Guidelines\_for\_Authors}, Dec. 2023.
\\[.5mm]\indent\ \copyright Institute of Computing Technology, Chinese Academy of Sciences 2021}

\noindent {\small\bf Abstract} \quad  {\small Neural network models for audio tasks, such as automatic speech recognition (ASR) and acoustic scene classification (ASC), are susceptible to noise contamination for real-life applications. 
To improve audio quality, an enhancement module, which can be developed independently, is explicitly used at the front-end of the target audio applications.
In this paper, we present an end-to-end learning solution to jointly optimise the models for audio enhancement (AE) and the subsequent applications.
To guide the optimisation of the AE module towards a target application, and especially to overcome difficult samples, we make use of the sample-wise performance measure as an indication of sample importance.
In experiments, we consider four representative applications to evaluate our training paradigm, i.e., ASR, speech command recognition
(SCR), speech emotion recognition (SER), and ASC.
These applications are associated with speech and non-speech tasks concerning semantic and non-semantic features, transient and global information,
and the experimental results indicate that our proposed approach can considerably boost the noise robustness of the models, especially at low signal-to-noise ratios (SNRs), for a wide range of computer audition tasks in everyday-life noisy environments.
}

\vspace*{3mm}


\vspace*{4mm}

\end{CJK*}
\baselineskip=15.8pt plus.2pt minus.2pt
\parskip=0pt plus.2pt minus0.2pt
\begin{multicols}{2}
\section{Introduction}
\label{sec:intro}
\Ac{CA} is one of the most prominent fields currently being revolutionised by the advent of \ac{DL}, with \acp{DNN} increasingly becoming the state-of-the-art in a multitude of applications, such as the ones discussed in this work: \ac{SCR}~\citep{de2018neural}, \ac{ASR}~\citep{baevski2020wav2vec}, \ac{SER}~\citep{Wagner22-DAWN}, and \ac{ASC}~\citep{ren2019attention}.
However, these applications are susceptible to different heterogeneities present in real-life conditions. 
Taking \ac{ASR} as an example, this may include amongst other within- and cross-speaker variations, for instance, disfluencies, differences in language, and recording devices and setups. 

One of the most prominent causes that impedes the practical application of \ac{CA} models is the innumerable types of ambient noises or interference that deteriorate the audio recording, including environmental background noise, interfering speakers, reverberation, etc. 
The sound of these noises can be stationary or non-stationary, instantaneous or continuous, and can be of different intensities (stable or variable), all of which have audio models confronting diverse and very complex situations. 
Meanwhile, in practical applications, multiple interfering sources can be present at the same time, each affecting the effectiveness of such audio models to a different extent.
Hence, while considerable performance improvements are leading to the continuous adoption of \ac{CA} modules in several \ac{AI} pipelines, \emph{robustness to noise} remains a critical consideration for most of them.
This has led to an accompanying rise of \ac{AE} methods, which typically also fall under the auspices of \ac{DL}~\citep{liu2021n}.

To that end, we present a novel framework for general \acl{AE} targeted towards increased robustness of different \acl{CAT}s.
In this framework, the cascaded \ac{AE} and \ac{CA} models perform two iterative training steps, strengthening the interplay between the different components of a \ac{DL} pipeline to minimise potential mismatches and benefit from potential synergies.
The motivation is that the \ac{CAT} model can guide the \ac{AE} frontend to preserve those signal components that are particularly important for the task at hand; for instance, an \ac{AE} frontend for \ac{ASR} might be optimised to improve the intelligibility of the signal, whilst a \ac{SER} frontend might focus on the preservation of prosody instead, as this property is more important for the identification of emotional information. 
Contrary to conventional joint optimisation, samples are not treated equally, but we utilise the loss of the target CAT model as an indication of difficulty in order to guide the training of \ac{AE} towards harder samples.

We hypothesise that the proposed training framework utilises the symbiotic and interdependent nature between the AE and CAT models, and thus counterbalances and mutually promotes the models to reach an optimal performance of the entire system.
The technique is experimentally assessed using four relevant target \ac{CA} applications, aiming to cover a broad spectrum from linguistic speech content in the case of \acl{ASR} and \acl{SCR}, to acoustic speech content in the case of \acl{SER}, to ambient audio in the case of \acl{ASC}.

The remainder of the paper is organised as follows. In \Cref{sec:method}, we provide an overview of our methodology, including the U-Net-based SE model, as well as the different CAT models. Then, we detail the utilised datasets and experiments and report our results in \Cref{sec:Experiments}, before putting said results in perspective in \Cref{sec:Discussion}. Finally, we conclude our work and point towards future research directories in \Cref{sec:Conclusions}.   

\begin{figure*}[ht!]
  \centering
    \begin{subfigure}[a]{0.25\textwidth}
        \centering
        \includegraphics[height=4in]{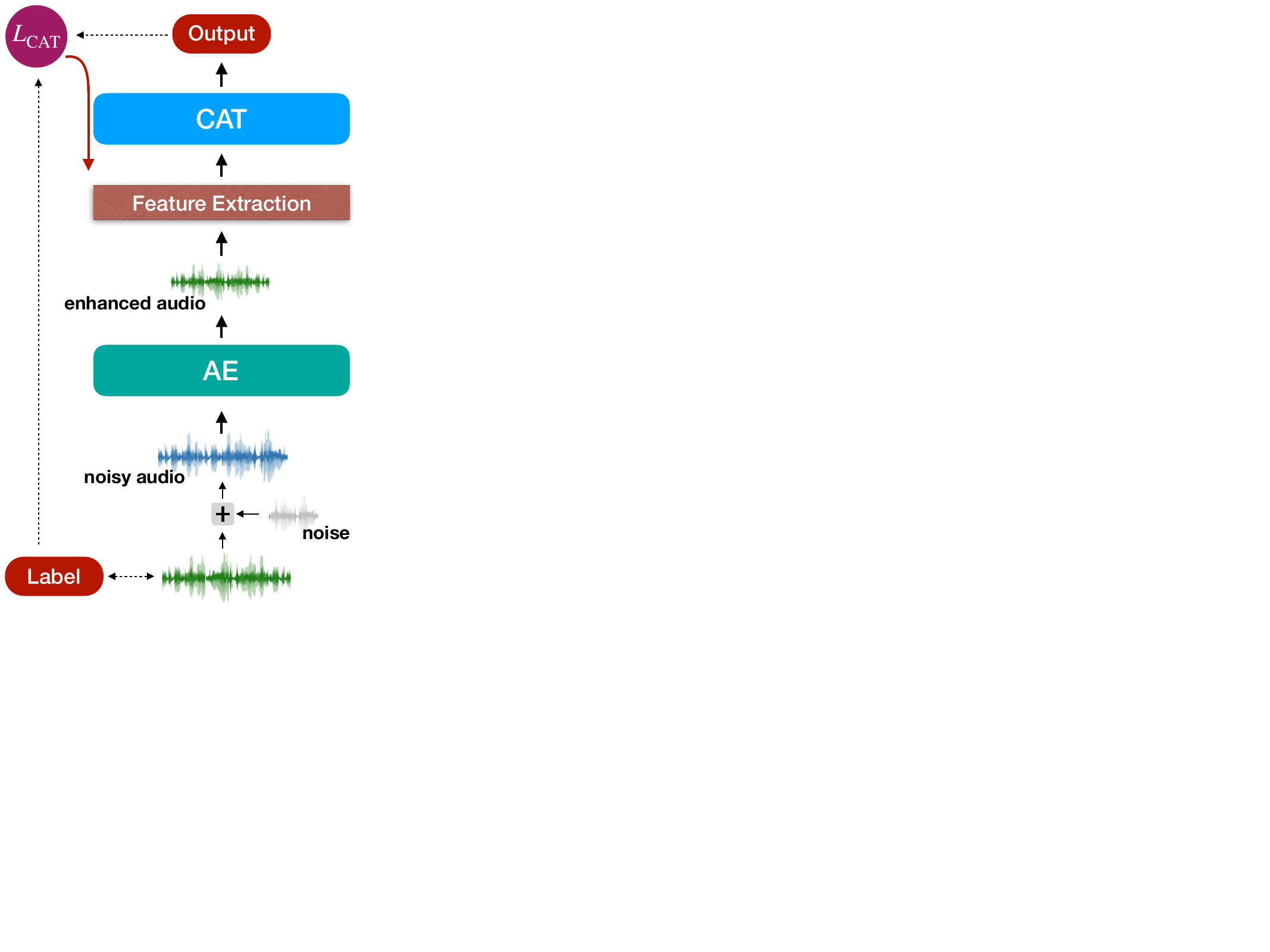}
        \vspace{-3.7cm}
        \caption{Cold cascade + data augmentation}
    \end{subfigure}
    ~
    \hspace*{1.5cm}
    \begin{subfigure}[a]{0.25\textwidth}
        \centering
        \includegraphics[height=4in]{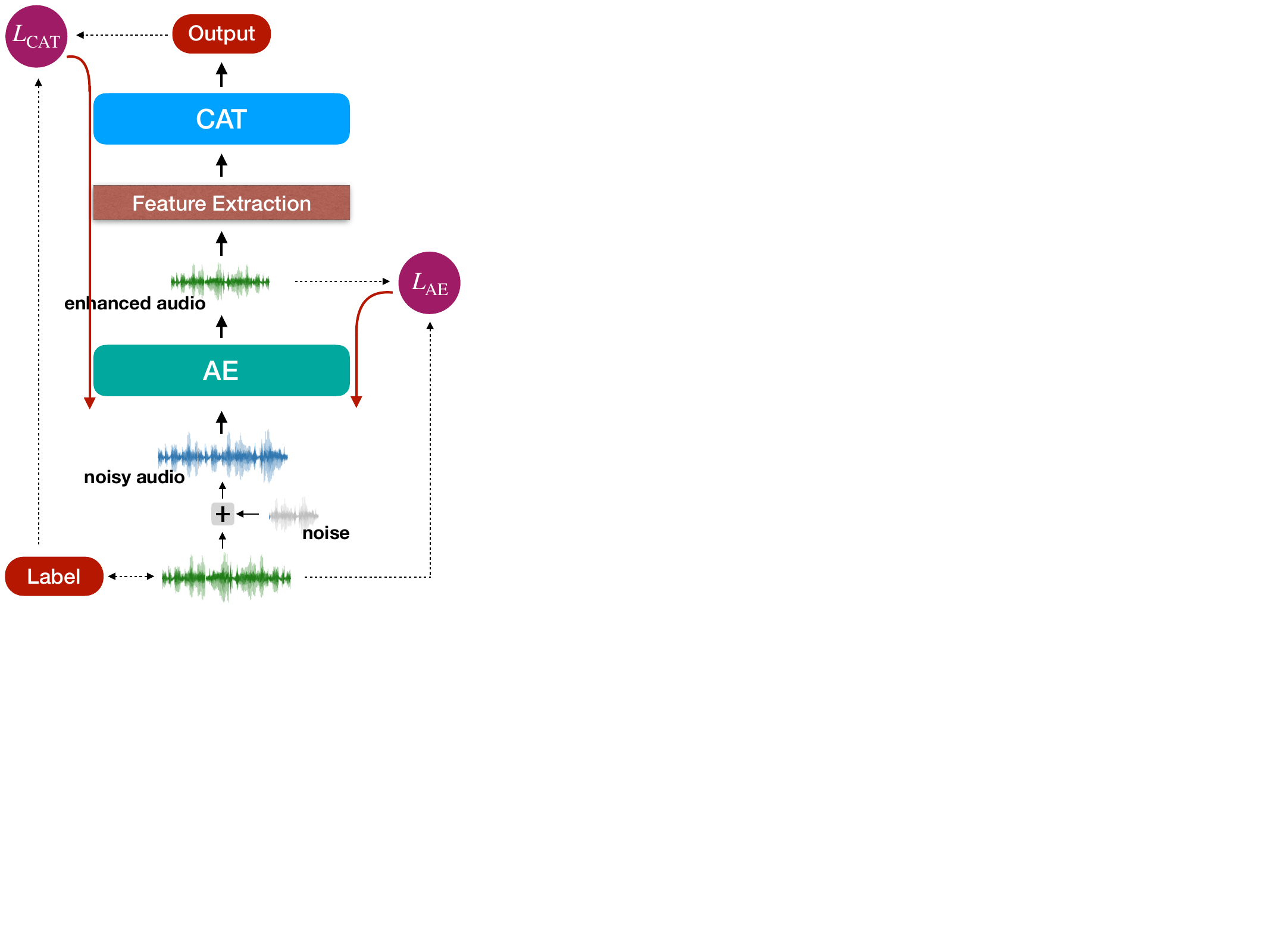}
        \vspace{-3.7cm}
        \caption{Multi-task learning}
    \end{subfigure}
    ~
    \hspace*{1.5cm}
    \begin{subfigure}[c]{0.25\textwidth}
        \centering
        \includegraphics[height=4in]{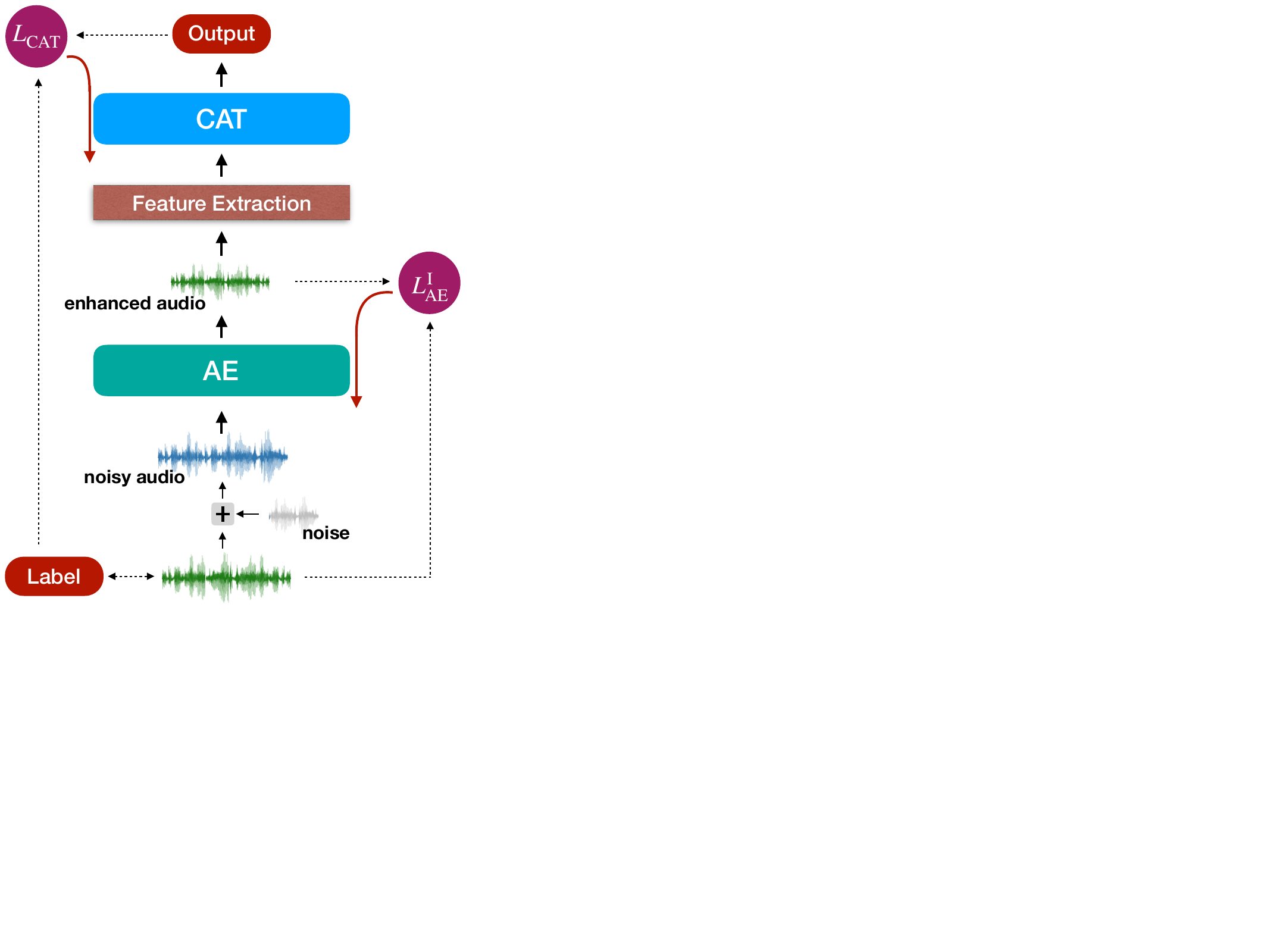}
        \vspace{-3.7cm}
        \caption{Iterative optimisation}
    \end{subfigure}
  \caption{Diagrams showing the methodologies used. The red arrows demonstrate the back-propagation through the network modules with respect to the losses $L$ of the AE and the CAT. In a) only the CAT loss is optimised with a frozen AE, whilst the optimisation in b) is based on the CAT and the AE loss with the AE parameters being affected through both losses. In our suggested approach c) the parameters of the CAT and the AE are only affected through their respective loss with the AE including a sample-level importance in contrast to the previous approaches.\label{fig:iterative} 
  }
\end{figure*}

\section{Related Work}
At its core, the task of \ac{AE} aims at the separation of the audio of interest from other interfering sounds, \ie, it aims at the preservation of the target signal while reducing the uncertainties in audio. The unwanted interference can be the result of several phenomena which affect different steps of the typical \ac{CA} pipeline:
\begin{enumerate*}[label=\alph*)]
    \item additive noise,
    \item reverberation,
    \item encoding noise, and,
    \item package loss.
\end{enumerate*}
From these, additive noise has been most thoroughly studied in previous work, due to its ubiquitous presence in \ac{CA} applications and its detrimental effects on performance~\citep{spille2018comparing, triantafyllopoulos2019towards}.

Within \ac{AE}, particular attention has traditionally been paid to \ac{SE}, as a typical \ac{CAT} is mostly focused on extracting information from the human voice.
\Ac{ASR}, being the flagship task of \acl{CA}, is the primary testbed for most \ac{SE} methods, with other tasks such as \ac{SER} and \ac{SCR} following closely. 
However, enhancement of audio signals beyond speech is needed in a number of  \ac{CAT}s, such as \ac{ASC} and \ac{SED}. 
In contrast to the purpose of speech enhancement,
the presence of speech is often deemed as the noise that can considerably affect the identification of surrounding environments~\citep{liu2020towards}. To tackle this problem, voice suppression, as another type of audio enhancement task, has the goal to eliminate the human voice from ambient recordings. These contradicting definitions of target audio signal and confounding noise show that a single one-shoe-fits-all solution for \ac{AE} systems seems rather difficult to achieve.

Utilising \emph{enhancement frontends}, \ie, separately developed enhancement modules (typically based on \acp{DNN}), can enhance the input for the subsequent \ac{CA} models,
which can explicitly be empowered using \emph{data augmentation} techniques, such as SpecAugment~\citep{daniel2019} or additive noise~\citep{triantafyllopoulos2019towards} for their better robustness against expected perturbations.
This is typically performed for ASR tasks \cite{weninger2015speech,kinoshita2020improving,sivasankaran2015robust,zorilua2019investigation}. However, in practice, such independent enhancement can introduce unwanted distortions and artefacts \cite{iwamoto2022bad} in the enhanced audio, yielding limited improvements or even worsen the performance of cascaded ASR models. 
In order to improve the tolerance to these distortions, the ASR model can be trained based on the enhanced audio, which is sometimes referred to as joint training \cite{wang2016joint,narayananlarge}. 

When optimising the ASR, the parameters of its frontend SE model can be either frozen or trainable. In the case of trainable parameters, the loss of the ASR task backpropagates through the whole combined model, \ie, the cascade of \ac{SE} and \ac{ASR}. This leads to a parameter update of the \ac{SE} model based on the \ac{ASR} loss~\cite{wang2016joint}. 
However, to have no explicit restrictions on the \ac{SE} model during the training poses a risk to weaken or even corrupt the SE effect.  
Considering the training objectives of SE and ASR together, recent work \cite{ma2021multitask,chen2015speech} frames the task into a multi-task learning problem, where the losses of SE and ASR are typically added up resulting in a combined optimisation. Other noteworthy approaches include the exploitation of more advanced deep learning techniques, such as generative adversarial networks (GANs) for SE \cite{liu19_interspeech,li2021adversarial}, and self-supervised learning (SSL) for ASR \cite{zhu2022joint}. The combination of the two losses is commonly scaled by a dynamic factor, which gradually shifts the training focus between the AE and ASR task \cite{kim2021streaming}. 
Joint training has also primarily been used in \ac{SCR} or keyword spotting \cite{cambara2022tase,gu2019monaural}, however, it is rarely used in other \ac{CA} applications. For most cases, the \ac{SE} module is trained separately and then cascaded to the target CA models for noise reduction, as for \ac{SER} \cite{triantafyllopoulos2019towards,zhou2020using} and \ac{ASC} \cite{liu2020towards}.

\section{Methodologies}
\label{sec:method}
At the core of our methodology we put two hypotheses, which are already partly supported by the literature, but have not yet been validated on a wide range of applications: 1) the enhancement of audio signals, which contain highly relevant information for a given computer audition task, as part of a processing pipeline, can improve the performance on the target task, and 2) a training procedure, which optimises the audio enhancement and the \ac{CAT} jointly can specialise the audio enhancement module for task-specific signals and therefore lead to better performance on the \ac{CAT}.

In order to further explore the hypotheses mentioned above, we report a set of experiments, based on \ac{AE} models using a U-Net architecture, to explore several training paradigms, including the joint optimisation of the audio enhancement for the \ac{CAT}s: automatic speech recognition, speech command recognition, speech emotion recognition and acoustic scene classification. Despite the difference in implementation details, the general framework stays the same amongst the different types of applications and is further illustrated in Fig.~\ref{fig:iterative}. 
All data for AE and CATs is resampled to 16\,kHz.

\subsection{Comparison Methods}

To assess the performance of our proposed iterative optimisation approach, we compare it with a wide range of methods commonly applied in the context of audio enhancement, which will be introduced in the following.\\

\noindent
\textbf{Baseline}\\
The general baseline for all experiments is a CAT-specific model taken from related literature, which is not trained on any noise-specific data and does not use an AE component. The model is not specifically designed for robustness towards noise and we thus expect a noticeable performance drop-off when confronted with noisy data.\\

\noindent
\textbf{Data Augmentation}\\
In a first attempt to make the baseline model more robust, we train it on noise-augmented data. For this purpose, we artificially add noise with different SNR ratios to the mostly clean audio recordings. With data augmentation being one of the most common machine learning practices to increase robustness, we expect the model to perform better on the noisy test data, which was generated in the same manner as the train data. \\

\noindent
\textbf{Cold Cascade}\\
The simplest training paradigm with an AE component is a cold cascade of U-Net, as described in section~\ref{sec:UNet} and CAT-specific model. Cold cascade means in this context that both models are being optimised independently. First, the U-Net is trained to achieve a good AE performance, then, the CAT model is trained based on clean data and then stacked on top of the U-Net.\\

\noindent
\textbf{Cold Cascade + Data Augmentation}\\
We further combine the cold cascade and data augmentation approach, \ie, first, the U-Net is trained to achieve a good AE performance, and then, 
we train the cold cascade architecture with augmented, noisy data. This approach promises decent noise robustness, as the model has previously seen noisy data and it includes a powerful AE component.\\

\noindent
\textbf{State-of-the-art}\\
To further evaluate the effectiveness of our methods against the state-of-the-art, we additionally utilise two recent denoising methods; this we only do for one of the CA tasks (SCR) due to space limitations.
Specifically, we use MetricGAN+~\citep{Fu21-MAI} and DeepFilterNet-3 (DFNet-3)~\citep{Schroeter23-DPM}.
MetricGAN+ is a bLSTM model trained in generative-adversarial fashion to optimise perceptual losses; the training set is VoiceBank-DEMAND~\citep{Valentini16-IRC}.
DFNet-3, on the other hand, follows a two-stage approach with ERB-based enhancement followed by deep filtering to enhance the periodicity of the output signal and has been trained with a multi-spectral loss on DNS-4~\citep{Dubey22-I2D}, which is closer to our current setup (\ie, the noise data partially comes from AudioSet). 
Both models are used to enhance the noisy mixtures of SCR on which we evaluate the baseline model trained without data augmentation on the original data; they thus simulate the scenario of using an off-the-shelf denoising model before evaluation.
This setup is essentially equivalent to Cold Cascade, only this time using different models.
\\

\noindent
\textbf{Multi-Task Learning}\\
We finally compare our method to an implementation of multi-task learning, \ie, an optimisation of both the \ac{AE} task and the \ac{CAT} at the same time with an additive loss function
\begin{equation}
    L = L_\mathrm{AE} + L_\mathrm{CA},
\end{equation}
where $L_\mathrm{AE}$ is the loss of the speech enhancement task as presented in \eqref{eq:lossAE} and $L_\mathrm{CA}$ is the loss of the computer audition task.

In contrast to common applications of multi-task learning, the two models do not only share a certain set of layers but the AE and CA models are put in sequence of each other, \ie, the AE loss is derived from an intermediate layer of the overall system.
Thus, minimising the \ac{AE} loss has no effect on the parameters of the \ac{CA} model, while the \ac{CA} loss back-propagates through the AE model.
Consequently, the \ac{AE} and the \ac{CA} losses, whilst working as mutual regularisation terms, introduce a bias towards the update of the AE parameters.
Similar ideas have been explored for the structure of a supervised auto-encoder \cite{NIPS2018_7296}.

\subsection{Iterative Optimisation}
Similar to the concept of multitask learning, the main motivation behind an iterative optimisation approach is a joint view of the two models. At its core, there are two hypotheses: 1) The CA model should always be adapted to the output of the AE model, which might contain  residual noise, introduced speech distortions, artefacts, etc. This specialisation to specific characteristics of the AE can be considered as a form of domain adaptation of the CAT, which has long been shown to help alleviate performance\cite{ben2006analysis} 2) the performance of the CAT can be utilised to move the focus of the AE model onto particularly difficult samples. Thereby, we aim at achieving the optimum result of the entire neural system, \ie, the front-end audio processing and the subsequent target applications.

The implementation of the iterative optimisation approach is straightforward, yet rarely considered in the literature: first, the optimisation of the AE model involves the loss of the target \ac{CAT} as a reference to indicate the difficulty of each training sample. This plays the role of sample-level importance to assist the \ac{AE} component to be biased towards relatively harder samples, for example, those corrupted by more intensive noise. Second, during training for the \ac{CAT}, the model should process the enhanced audio signal, rather than the completely clean signal, in order to avoid a common performance gap resulting from a cold cascade of the front- and back-end models. 
However, as long as the \ac{AE} model is optimised, a more robust CA model needs to be adapted to the enhanced audio. On the other side, a more robust CA model can further assist the optimisation of the AE model by updating the difficulties of new samples.
Having this in mind, we hypothesise that both optimisation steps need to be performed iteratively to gradually approach an optimal solution.

In order to implement the latter idea, we calculate a weight for each sample in a given batch when optimising the AE model. The weight for each sample $i$ is defined as 
\begin{equation}
\label{eq:weightsCA}
    w_i = L_\mathrm{CA}(t_i, \hat{t}_i),
\end{equation}
with the target $t_i$ and the predicted target $\hat{t}_i$.
The weights therefore give an indication of how difficult a given sample is for the \ac{CAT}. We choose a linear relationship between the sample weight and the loss for the CAT as the most straightforward implementation, even though other approaches, such as softmax normalisation, are possible. This choice does not add any new hyperparameters, as linear scaling would only affect the training in the same way as changing the learning rate. In practice, we normalise the weights by dividing by their sum within one batch. 
The loss of the \ac{AE} component is then defined as 
\begin{equation}
\label{eq:SEiterativeloss}
    L_{AE}^\mathrm{I}(x, \hat{x}) = \frac{1}{N} \sum_{i=1}^{N} w_i L_{AE}(x_i, \hat{x}_i),
\end{equation}
with the noisy inputs $x_i$ and the reconstructed signals $\hat{x}_i$.
In the iterative training paradigm, we alternate with each batch by first optimising the CA system based on the AE output, while freezing the parameters of the AE system, and secondly optimising  the AE system according to the loss \eqref{eq:SEiterativeloss}, while freezing the parameters of the CA model in the weights \eqref{eq:weightsCA}.
The iterative training augments the interplay between the two models by persistently adapting the CA to the improved SE model while the updated CA can further be used as an indicator to improve the SE model.

\subsection{Audio Enhancement Model}
\label{sec:UNet}
\begin{figure*}[ht!]
  \centering
    \includegraphics[scale=0.4]{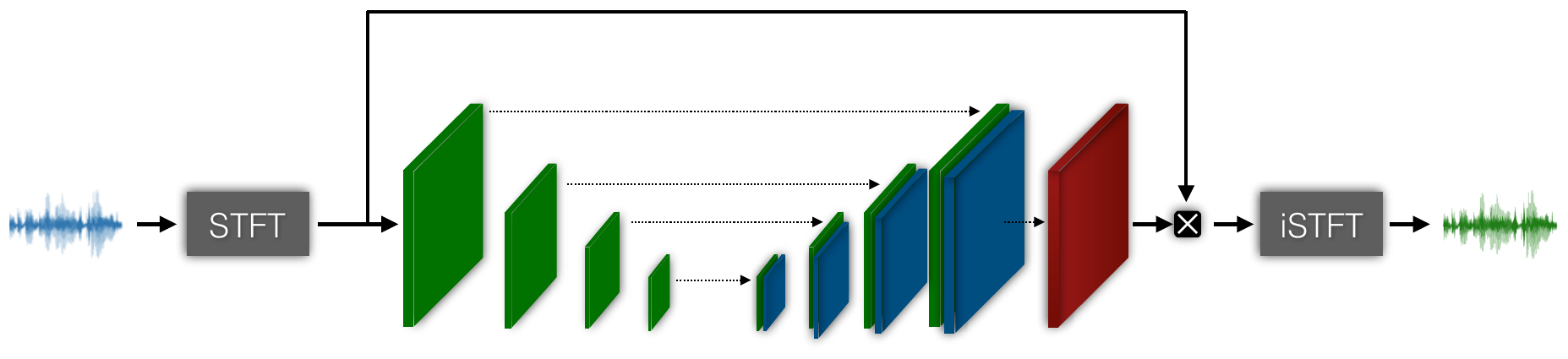}
  \caption{Schematic diagram of the U-net architecture. The raw audio is transformed with a short-time Fourier transform (STFT) into a spectrogram, which is then fed into a fully convolutional network with an encoder and decoder and skip connections between corresponding encoder and decoder layers in the U-shaped architecture. The final reconstructed or enhanced spectrogram is then transformed back into a raw audio signal with an inverse STFT.}\label{fig:Unet}
\end{figure*}

The audio enhancement is based on U-net~\cite{Ronneberger2015UNetCN,choi2018phase}, an auto-encoder architecture, operating in the frequency domain, with feed-forward layers that stack the encoder layers to their corresponding decoder layers, as seen in Fig.~\ref{fig:Unet}.

\begin{figure*}[ht!]
  \centering
    \begin{subfigure}[a]{0.20\textwidth}
        \centering
        \includegraphics[height=1.5in]{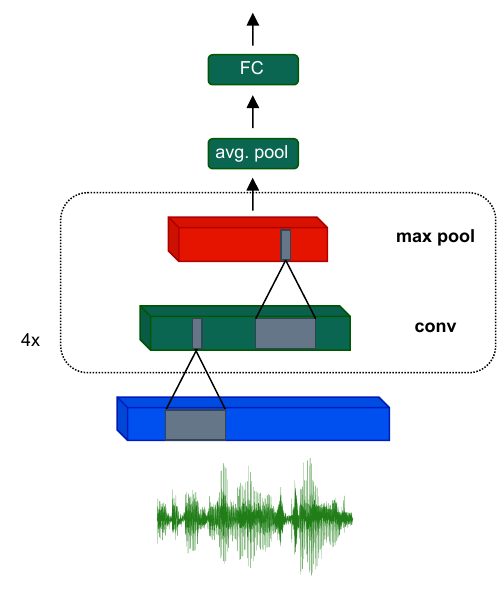}
        \caption{\centering{Speech Command Recognition}}
        \label{fig:models(a)}
    \end{subfigure}
    {}
    \begin{subfigure}[a]{0.2\textwidth}
        \centering
        \includegraphics[height=1.5in]{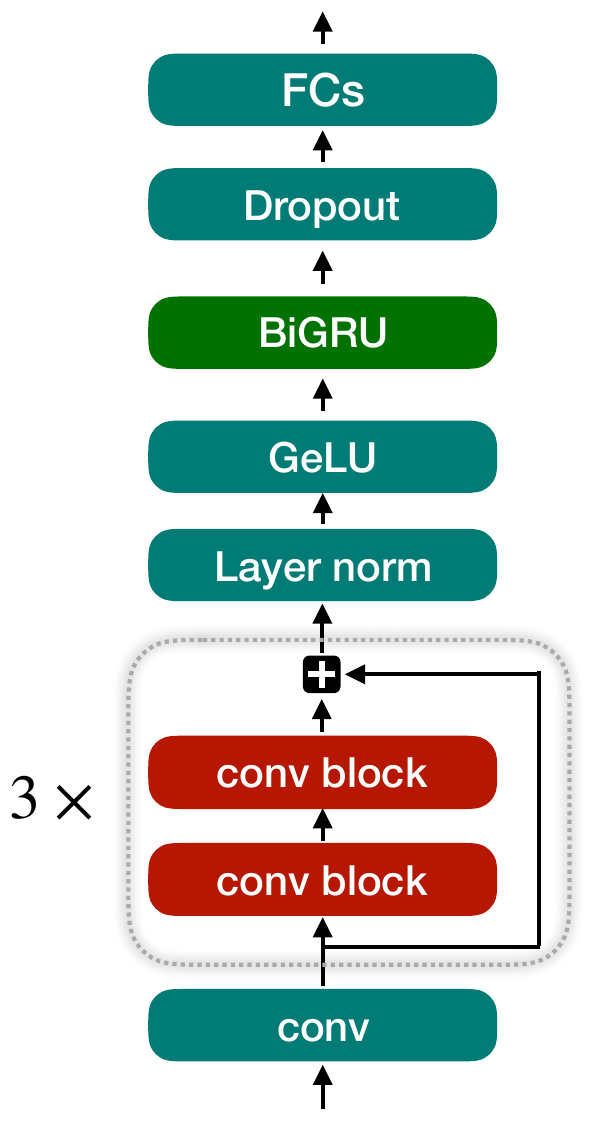}
        \caption{\centering{Automatic Speech Recognition}}
        \label{fig:models(b)}
    \end{subfigure}
    {}
    \begin{subfigure}[c]{0.2\textwidth}
        \centering
        \includegraphics[height=1.5in]{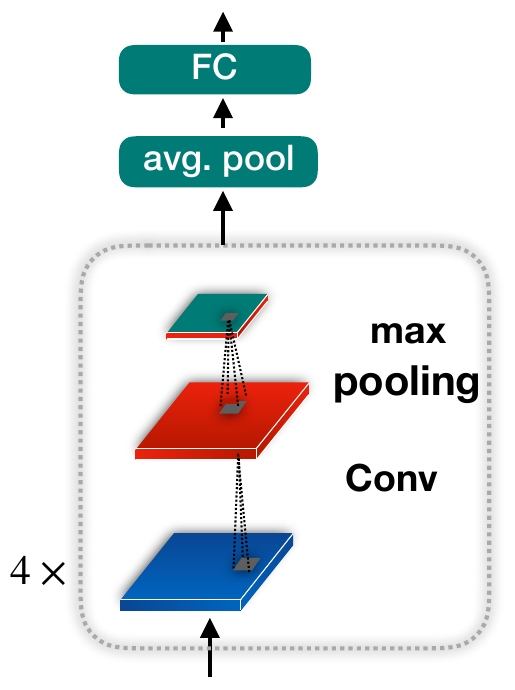}
        \caption{\centering{Speech Emotion Recognition}}
        \label{fig:models(c)}
    \end{subfigure}
    \begin{subfigure}[c]{0.2\textwidth}
        \centering
        \includegraphics[height=1.5in]{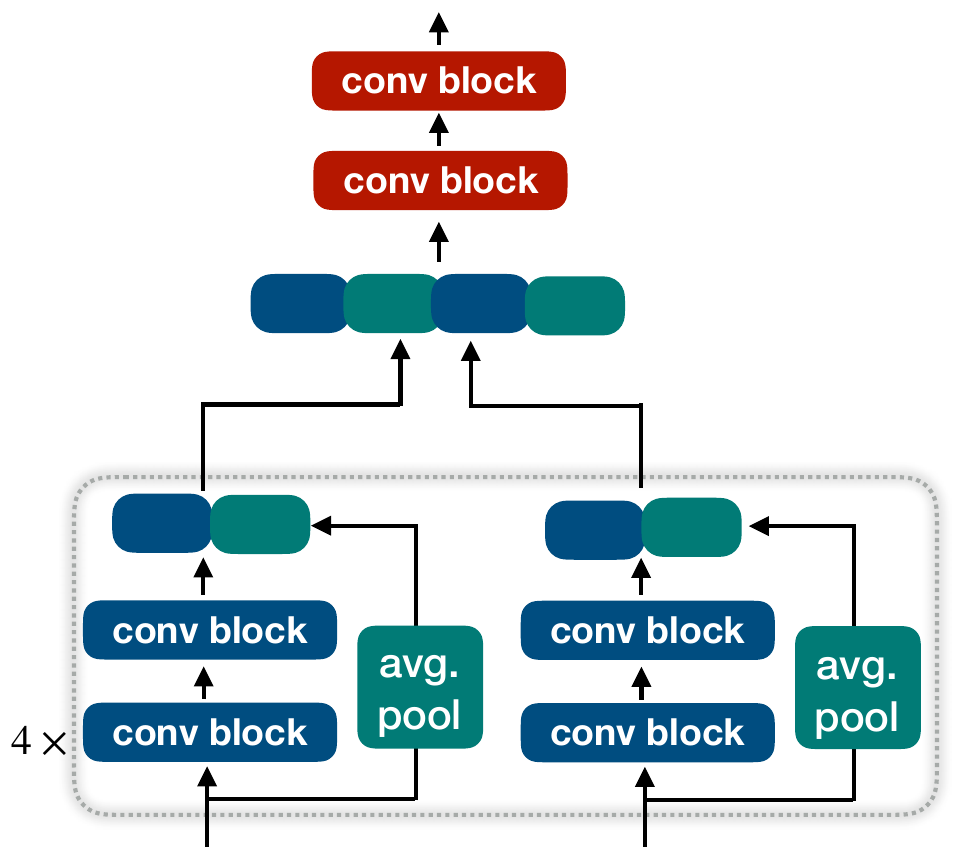}
        \caption{\centering{Audio Scene Classification}}
        \label{fig:models(d)}
    \end{subfigure}
  \caption{Schematic diagrams showing architectures for downstream computer audition tasks. The architecture for speech command recognition in a) is the only one acting on the raw audio signal compared to the architectures of b) to d), which take 2-dimensional spectrogram representations of the audio signal as an input. In a) we apply 1D convolutional and maxpooling layers, prior to a global average pooling and classification layer. The automatic speech emotion recognition model depicted in b) consists of 2D convolutional layers and convolutional blocks with skip connections, followed by a layer normalisation and a bi-directional GRU-RNN layer prior to the classification layer. The Speech emotion recognition architecture in c) only applies convolutional blocks prior to a global average pooling and a classification layer. The audio scene classification model in d) concatenates the outputs of different convolutional blocks and works in a fully convolutional manner.}\label{fig:models_specific} 
\end{figure*}

Given a noisy audio $y$ and its corresponding clean sample $x$, the noisy sample is converted into a spectrogram $Y$, using the short-time Fourier transform (STFT). The U-Net estimates a ratio mask $\operatorname{Mask}(Y)$, which is then applied to the original noisy input to predict the clean spectrogram:
\begin{equation}
    \hat{X} = Y \cdot \operatorname{Mask}(Y).
\end{equation}
The estimated clean audio $\hat{x}$ can then be reconstructed by applying inverse STFT. The parameters of the model are optimised by minimising the weighted SDR (wSDR) loss of the original and the estimated clean speech and noise \cite{choi2018phase}:
\begin{equation}
\label{eq:lossAE}
    L_{AE}(x, \hat{x}) = \alpha L_{SDR}(x, \hat{x}) + (1-\alpha) L_{SDR} (n, \hat{n} ),
\end{equation}
where 
\begin{equation*}
    n = y - x 
    \quad 
    \mathrm{and}
    \quad 
    \hat{n} = y - \hat{x}
\end{equation*}
represent the true and estimated noise signal, and 
\begin{equation}
L_{SDR}(x, \hat{x}) = - \frac{<x, \hat{x}>}{\vert\vert x\vert\vert\cdot \vert\vert\hat{x}\vert\vert}, 
\end{equation}
as well as
\begin{equation}
    \alpha = \frac{\vert\vert x\vert\vert ^2}{\vert\vert x\vert\vert^2+\vert\vert n\vert\vert ^2}. 
\end{equation}

In order to capture the advantages of enhanced audio signals from U-Net some slight architectural changes need to be applied in order to make it compatible in a cascading fashion with any of the above-mentioned application scenarios. For this purpose, we set the max-pooling along the time-axis equal to 1, while the pooling along the frequency-axis stays unchanged. The main motivation of this step is to allow the U-Net to process audio segments of different lengths, which is a crucial ability for some of the application tasks, like for instance ASR.

Any of the considered \ac{CAT} can do further processing, like feature extraction based on the enhanced waveform, as it would normally be done on the original waveform. Alternatively, the reconstructed time-frequency features of the AE model can directly be passed on. In a cold cascade, the AE model is first optimised based on its loss $L_\mathrm{AE}$ according to \eqref{eq:lossAE} in order to obtain a decent AE model for pre-processing before the CA model is trained for its task independently.
Beyond the U-Net architecture, we further investigated a complex U-Net\cite{choi2018phase} architecture, as well as a Wave U-Net architecture\cite{stoller2018wave}. An analysis with respect to Cepstral Distortion (CD), signal-to-distortion ratio (SDR), short-time objective intelligibility (STOI) and log spectral distortion (LSD) however showed a superior performance of the U-Net compared to the other candidate models. Experiments for downstream CATs where therefore only carried out with the U-Net architecture in order to limit the already computation-heavy experiments.

\subsection{Computer Audition Tasks}
In the following, we will introduce the four different computer audition tasks namely \textit{speech command recognition} (SCR), \textit{automatic speech recognition} (ASR), \textit{speech emotion recognition} (SER) and \textit{audio scene classification} (ASC), as well as the corresponding NN architectures, on which we evaluated our iterative training strategies. An overview of the applied architectures is given in Fig. \ref{fig:models_specific}.

\subsubsection{Speech Command Recognition}
SCR belongs to the category of tasks, in which linguistic information has to be extracted from speech. 
Common SCR tasks are implemented such that audio recordings, potentially of identical length, have to be assigned to one speech command or a single word out of a given set of commands or vocabulary. 
Due to the limitations regarding the variability of audio and labels, SCR can be considered less complex compared to general ASR and solutions do not necessarily contain language models.
Hence, models for SCR have the potential to be designed shallowly in order to run on mobile edge devices or other assistive devices without the necessity of an internet connection~\citep{de2018neural}.

We evaluate our methodology on the 35-word limited-vocabulary speech recognition data set introduced in \cite{sc2018}. Accordingly, we choose the M5 version of the very deep CNN, as introduced in \cite{dai2017very} with 35 neurons in the softmax output layer. The network consists of a set of 1D convolutional layers acting on the raw waveform in its time-domain without further pre-processing. Fig~\ref{fig:models(a)} provides a visualisation of the approach.

\subsubsection{Automatic Speech Recognition}

ASR is certainly one of the most prominently researched problems in CA, as the automatic transcriptions of spoken language 
have a multitude of applications, which are already available in commercial devices. As the nature of speech can be considered quite complex, general ASR models need to cope with different speaker characteristics, such as different speaking speeds, and, in general, a variable length of sentences. 
Applications of ASR are manifold and are a cornerstone of human-machine interaction (HMI), for instance in digital assistants, such as Alexa, Siri, and alike. 
The era of deep learning has helped boost the performance of ASR systems, which have previously been dominated by Hidden Markov Models-Gaussian Mixture Models (HMM-GMM)~\cite{wang2019overview}.

Common architectures for ASR tasks can be split into two components: an acoustic model, which finds a probability-based mapping between spoken utterances and characters within an alphabet, and a language model, which converts the probability distribution to coherent text. Most state-of-the-art acoustic models are based on self-supervised learning (SSL), which can be employed to learn powerful representations from large-scale data, which has not previously been annotated. The learnt representations find application beyond ASR \cite{baevski2020wav2vec,hsu2021hubert,babu2021xlsr} in multiple
downstream tasks \cite{jing2020self,liu2021self}.

In order to explore the idea of iterative optimisation however, we choose an acoustic model, which is not relying on SSL, as an SSL-based system would not be compatible with our training paradigm. Instead, we choose an architecture similar to Deep Speech 2~\cite{pmlr-v48-amodei16}. 
The basis of the architecture is a set of three residual blocks, each of which consists of two convolutional layers with batch normalisation, GeLU activation, and dropout. The addition of skip-connections in our model compared to Deep Speech 2, promises a more stable convergence~\cite{li2018,zheng2020}. The output of the residual blocks is then further processed by several recurrent layers with bidirectional gated recurrent units (GRUs) leading to speech representations, which capture the temporal dynamics in the speech signal. The final layer of the architecture is a fully connected layer, which maps the input to the character indices of the alphabet. In order to improve the quality of the recognised text from the acoustic model,
we apply beam search with a 3-gram ARPA language model on the output of the acoustic model. An overview of the system is depicted in Fig.~\ref{fig:models(b)}\footnote{The language model can be found at \href{https://www.openslr.org/11/}{https://www.openslr.org/11/}.}.

Given the importance of ASR in research, there have already been an extensive amount of studies investigating the robustness of ASR models against noise. Common approaches utilise data augmentation techniques, either by distorting the frequency-domain representation of audio, like in the case of SpecAugment~\cite{daniel2019}, by incorporating additive synthesised noise to the clean speech samples~\cite{hannun2014deep,yin2015}, or in a teacher-student architecture, in which the student network is gradually taught to adapt to noise~\cite{kim2018,meng2019domain}.
For the input to the network, we extract Mel spectrograms from the raw audio using STFT applying a step size of 10 ms and a window length of 20 ms with 32 Mel-scale filters.

\subsubsection{Speech Emotion Recognition}
\Acl{SER} is a cornerstone technology for the development of successful HMI applications~\citep{Schuller18-SER}.
It involves the development of algorithms that can understand human emotions from vocalisations and is typically formulated as a classification (of `basic' emotions) or a regression task (of emotional dimensions)~\citep{Schuller18-SER} and studies often focus on specific contexts, such as to recognise acted emotions~\cite{Busso2008IEMOCAPIE}, emotions in public speaking scenarios~\cite{baird21} or emotions of individuals with autism~\cite{milling22}.
While the field has seen tremendous progress in recent years, especially with the increasing improvement of \ac{DL} algorithms~\citep{Wagner22-DAWN}, \emph{robustness} remains a key issue.
In particular, \ac{SER} models have been shown to suffer from susceptibility to encoding errors~\citep{Oates19-RSE}, packet loss~\citep{Mohamed20-PLC}, and additive noise~\citep{triantafyllopoulos2019towards, Wagner22-DAWN}.
Of those, additive noise is the more insidious, as it is beyond the control of the application designer (unlike encoding errors and packet loss which can be fixed by other means) and needs to be addressed with \acl{AE} methods.

In recent years, \ac{SER} research has transitioned to the use of \ac{DL} models like \acp{CNN}~\citep{Triantafyllopoulos21-MLC}, an approach we follow here as well.
In particular, we use a 4-layered \ac{CNN}, where each layer consists of a sequence of convolution, batch normalisation, ReLU activation, max-pooling, and dropout.
Its input consists of the Mel spectrogram, computed with 32 Mel-scale filters, a window length of 20\,ms, and a step size of 10\,ms.
This architecture has been shown to be effective in previous works\todo{which ones?}. 
Its output is projected to emotion labels using a dense layer, as depicted in Fig. \ref{fig:models(c)}.

\subsubsection{Acoustic Scene Classification}
Our final audio application, \ac{ASC}, is concerned with the classification of soundscapes in discrete categories that characterise their content (\eg, a park or a shopping mall).
This application departs from the standard assumption that speech is the signal to be preserved.
Instead, speech is now considered a contaminating source which needs to be removed.
There are two primary motivating factors for this unorthodox formulation: a) improving the robustness of \ac{ASC} classification in the presence of speech~\citep{liu2020towards}, and b) enforcing privacy regulations in the case of large-scale, monitoring applications~\citep{bajovic2021marvel}.
In fact, the two factors have a strong overlap as data collection for \ac{ASC} applications typically takes mitigating steps to avoid the capturing of speech (\eg, filtering out segments where a VAD is triggered~\citep{bajovic2021marvel}) resulting in datasets that do not violate privacy requirements, but will have trouble generalising to real-world environments where human speech is ubiquitous.
To that end, we propose to enhance \ac{ASC} signals by removing speech -- a form of \emph{voice suppression}~\citep{liu2020towards}.

As our \ac{ASC} model, we use Dual-ResNet~\citep{9053274}, which was awarded as the most reproducible system for the first task of the 2020 \ac{DCASE} challenge~\citep{Heittola2020}. 
The model contains two different paths for separately processing the low- (lower 64) and high-frequency bands (upper 64). 
Late fusion is used to concatenate the outputs of these two paths, before going through two additional $1\times1$ convolutional layers to reduce the dimensionality to the number of classes. 
A schematic visualisation of the architecture can be found in Fig.~ \ref{fig:models(d)}.
The low- and high-frequency paths have an identical architecture, namely, a residual network of 8 convolutional blocks, each block a sequence of batch normalisation, ReLU activation, and a final convolutional layer. 
The Log Mel spectrogram of $128$ Mel-bands, extracted from the audio waveform by applying STFT with the window length of 64\,ms and a hop size of 16\,ms, are used as the model input.\\

\subsection{Training Details}

During the training we applied a batch size of 16 for the U-Net audio enhancement, which has shown optimal performance in preliminary experiments.  
All models are trained with an Adam optimiser and additional weight decay is applied for the SCR and ASC models in the form of L2 regularisation. The ASR model is trained with a connectionist temporal classification (CTC) loss~\cite{graves2006connectionist}, whilst we optimise the cross-entropy loss for the remaining CATs. 
For the AE, ASR and ASC models, we set the learning rate to 0.0001, while we set it to 0.001 for SER. For the SCR task, we reduce an initial learning
rate of 0.01 to 0.001 after 20 epochs. In order to train the audios of varying lengths for CATs like ASR and SER we pad the shorter audios to the length of the longest sample.

\section{Experimental Results}
\label{sec:Experiments}
We conducted experiments to evaluate the proposed training paradigm
on all four \ac{CAT}s introduced. In the following, we describe the datasets, evaluation metrics, and experimental setups for each case and report our results. 
\footnote{Examples to show our audio enhancement performance, including speech enhancement for different languages and voice suppression, can be found in: \url{https://github.com/EIHW/AE_SampleImportance}} 
\begin{table*}[ht!]
  \begin{center}
  \caption{Speech command recognition testing results,  (Acc)uray[$\%$], using the Speech commands data set and the AudioSet corpus. DA stands for the method using only data augmentation. MTL represents the proposed multi-task learning solution.}
  \renewcommand{\arraystretch}{1.2}
  \begin{tabular}{l c c c c c c c c}
    \toprule
    Methods & Inf & $25\si{\deci\bel}$ & $20\si{\deci\bel}$ & $15\si{\deci\bel}$ & $10\si{\deci\bel}$ & $5\si{\deci\bel}$ & $0\si{\deci\bel}$ & average\\
    \midrule
    original SCR & $85.07$ & $83.37$ & $81.35$ & $76.87$ & $67.57$ & $51.52$ & $33.12$ & $65.63$\\
    DA & - & $82.69$ & $82.07$ & $80.09$ & $77.53$ & $71.66$ & $58.26$ & $75.38$\\
    Cold Cascade & - & $84.34$ & $83.38$ & $80.85$ & $75.54$ & $66.06$ & $51.92$ & $73.68$\\
    Cold Cascade + DA & $-$ & $82.65$ & $82.31$ & $81.64$ & $79.31$ & $74.22$ & $64.93$ & $77.51$\\
    MetricGAN+ & - & 77.78 & 75.81 & 71.95 & 64.49 & 52.60 & 35.37& 63.00 \\
    DFNet-3 & - & 80.81 & 79.83 & 78.07 & 74.89 & 69.72 & 62.88 & 74.37\\
    \hline
    \textbf{MTL} & - & $\textbf{85.53}$ & $\textbf{84.21}$ & $82.12$ & $80.04$ & $76.54$ & $67.18$ & $79.27$\\
    \textbf{iterative optimisation} & - & $85.35$ & $83.93$ & $\textbf{82.37}$ & $\textbf{81.56}$ & $\textbf{77.41}$ & $\textbf{69.18}$ & $\textbf{79.97}$\\
    \bottomrule
  \end{tabular}
  \label{Results_Table_SCR}
  \end{center}
\end{table*}

\subsection{Downstream Task I: Speech Command Recognition}
The first \ac{CAT} application investigated in this work is SCR based on the limited-vocabulary dataset
\cite{sc2018}. The data includes 105\,829 one-second-long audio clips of 35 common words, including digits zero to nine, fourteen words, which are considered useful as commands for IoT and robotics, as well as some additional words covering a variety of phonemes. The data further provide instances, which contain only background noise or speech that does include the target words. The negative samples are expected to help the keyword spotting systems to differentiate relevant from non-relevant audio clips, thereby lowering false positives in applications. 
The problem at hand is thus formulated as a 35-class classification task based on an audio recording of constant length. Given the quite balanced distribution amongst classes in the test set, we choose accuracy, \ie, the ratio of correctly classified samples over all samples, for evaluation. 

In order to apply our audio enhancement methodology, we augment the original (`clean') data with noise recordings from AudioSet, which are truncated to a length of one second. 
The two signals are added considering uniformly distributed SNR levels between $0$ and $25$\,\db, covering a range from very low volume noise (at 25\,\db) to equal volume of noise and target signal (0\,\db).
Since the applied SCR model directly acts on the raw audio, no further data processing is necessary.
We evaluated our model at the constant SNR levels $25, 20, 15, 10, 5$ and $0$\,\db. We picked these specific SNR levels to show how our models deal with increasingly noisy samples from the test set.
The AudioSet corpus contains more than two million human-labelled 10\,s environmental sound clips drawn from YouTube videos. 
We exclude all human-related noise samples labelled as ``human sounds'' from AudioSet and obtain 16\,198 samples for the training set, 636 samples for the development set, and 714 samples for the test set.
The model architecture --as described in \Cref{sec:method}
-- is independently trained for the different noise levels and the training paradigms, as described in \Cref{sec:method}, \ie,  \textit{baseline}, \textit{data augmentation}, \textit{cold cascade}, \textit{cold cascade + data augmentation}, \textit{multi-task learning}, and \textit{iterative optimisation}. 

The baseline of the SCR model without additive noise achieves an accuracy of $85.07\,\%$ (cf.\ Table \ref{Results_Table_SCR}). Intuitively, the performance of the same system decreases monotonically with increasing noise levels, dropping to $33.12\,\%$ with 0\,dB SNR. All of the suggested approaches aim at increased robustness to help mitigate said drop-off. This effect becomes more noticeable with lower SNR values as, at 0\,dB, even the worst improvement compared to the baseline alleviates the accuracy to more than $50\,\%$. The suggested iterative optimisation and MTL training paradigms outperform competing approaches in every instance, with the MTL achieving slightly better performance at high SNRs and the iterative optimisation performing better on low SNRs. At 0\,dB, the iterative optimisation allows for an accuracy more than twice as high as the baseline. Noticeably, at 25\,dB, MTL and iterative optimisation even outperform the baseline without additive noise. One possible explanation for this effect is that the AE filters small levels of inherent noise in the ``clean'' data itself. However, this claim is hard to verify, as quantitative measures of noise levels without completely noise-free ground-truths to compare against are difficult to obtain, making a deeper analysis necessary.

Our iterative optimisation and MTL methods also perform favourably with respect to the state-of-the-art.
DFNet-3 denoising achieves an average accuracy of $74.37,\%$, which is substantially lower than our $79.97\%$.
The same is true for MetricGAN+, which ranks lower even than the baseline model at $63.00\%$; this failure particularly illustrates how difficult the task is, and how a simple denoising frontend can fail.
Both models underperform the baseline at higher SNRs, which indicates that they introduce some unwanted distortion into the signal -- something that our methods avoid.
We also note that DFNet-3 is only marginally better than our own Cold Cascade method, even though the DNS-4 dataset is vastly bigger and more diverse than ours, which shows that our model is competitive in terms of enhancement performance.
Overall, the comparison to state-of-the-art illustrates that iterative optimisation is crucial for bridging the gap to downstream performance between clean and noisy audio.

\begin{table*}[ht!]
  \begin{center}
  \caption{\label{tab:asr_librispeech_results}Automatic speech recognition testing results, WER [$\%$], using Librispeech and the AudioSet corpus. DA stands for the method using only data augmentation. MTL represents the proposed multi-task learning solution.}
    \renewcommand{\arraystretch}{1.2}
  \begin{tabular}{l c c c c c c c c}
    \toprule
    Methods & Inf & $25\si{\deci\bel}$ & $20\si{\deci\bel}$ & $15\si{\deci\bel}$ & $10\si{\deci\bel}$ & $5\si{\deci\bel}$ & $0\si{\deci\bel}$ & average\\
    \midrule
    original ASR& $7.84$ & $10.74$ & $13.53$ & $19.87$ & $31.97$ & $49.72$ & $68.46$ & $32.38$\\
    DA & - & $9.58$ & $10.18$ & $11.17$ & $14.50$ & $21.05$ & $35.46$ & $16.99$\\
    Cold Cascade & - & $9.53$ & $10.84$ & $13.31$ & $18.16$ & $28.07$ & $43.78$ & $20.62$\\
    Cold Cascade + DA & $-$ & $8.15$ & $8.76$ & $10.03$ & $13.30$ & $20.89$ & $34.67$ & $15.97$\\
    \hline
    \textbf{MTL} & - & $\textbf{8.03}$ & $\textbf{8.69}$ & $\textbf{9.91}$ & $12.93$ & $19.45$ & $32.64$ & $15.27$\\
    \textbf{iterative optimisation} & - & $8.35$ & $8.79$ & $10.00$ & $\textbf{12.71}$ & $\textbf{19.27}$ & $\textbf{31.93}$ & $\textbf{15.18}$\\
    \bottomrule
  \end{tabular}
  \label{Results_Table_SE}
  \end{center}
\end{table*}

\begin{table*}[ht!]
\setlength{\tabcolsep}{7.5pt}
  \begin{center}
  \caption{Automatic speech recognition testing results, WER [$\%$], using the CHiME-4 challenge set. DA stands for the method using only data augmentation. MTL represents the proposed multi-task learning solution.}
  \renewcommand{\arraystretch}{1.2}
  \begin{tabular}{l c c c c c c }
    \toprule
    & \multicolumn{2}{c}{\textbf{GMM-HMM}} & \multicolumn{2}{c}{\textbf{DNN-HMM}} \\
    \cmidrule(lr){2-3}
    \cmidrule(lr){4-5}
    \textbf{Method} & \textbf{simu} & \textbf{real} & \textbf{simu} & \textbf{real}\\
    \midrule
    original ASR & $24.46$ & $22.19$ & $12.96$ & $11.56$\\
    Cold Cascade 1 & $18.48$ & $18.06$ & $12.54$ & $11.14$\\
    Cold Cascade 2 & $16.06$ & $14.59$ & $11.15$ & $9.50$\\
    \hline
    \textbf{MTL} & $15.04$ & $12.76$ & $9.88$ & $8.73$\\
    \textbf{iterative optimisation} & $\mathbf{14.08}$ & $\mathbf{12.53}$ & $\mathbf{9.45}$ & $\mathbf{8.12}$\\
    \bottomrule
  \end{tabular}
  \label{tab:CHiME}
  \end{center}
\end{table*}

\subsection{Downstream Task II: Automatic Speech Recognition}
ASR experiments are performed on two datasets, the first of which being Librispeech~\cite{panayotov2015librispeech}, as for the previous task noise-enhanced with AudioSet recordings, the second of which being the already artificially noise-enhanced dataset of the CHiME-4 challenge.
LibriSpeech consists of approximately 960\,hours of read, clean speech derived from over 8\,000 public domain audiobooks, containing its own train, development, and test splits. The data set has previously been used for similar tasks under the assumption of not containing any noise contamination \cite{liu2021n, liu2023towards}. The CHiME dataset is based on the Wall Street Journal (WSJ0) corpus. The training set mixes clean speech with noisy backgrounds, leading to 35\,690 utterances from 83 speakers in 4 different noisy environments. The test set contains simulated recordings and utterances recorded in real-world noisy environments from 4 other speakers.

We choose Mel-frequency cepstral coefficients (MFCCs) as input features for the ASR model with 40 Mel-band filters. The MFCCs are based on the STFT and a mapping onto the Mel scale with triangular overlapping windows, followed by a discrete cosine transform. Note that for the MTL and the iterative optimisation approach, the MFCCs are extracted from the output of the AE.
For evaluation of the ASR task, we use Word Error Rate (WER), an evaluation metric measuring  misclassification rate with respect to the words in an utterance, which is commonly used in the literature and therefore allows for a reasonable
comparisons to previous works.

The results of our experiments on Librispeech are summarised in Table~\ref{tab:asr_librispeech_results}. 
Naturally, the best result is obtained with the noise-free samples at a WER below 8. Even though this result, as well as the benchmark architecture are below state-of-the-art, we expect the benefits of our SE approach to translate to state-of-the-art models.
Similar to the SCR case, we observe that the iterative optimisation outperforms all other approaches for the low SNRs (10\,dB and lower), whilst the MTL shows the best performance for high SNRs (15\,dB and higher) with the DA and cold cascade + DA approaches performing slightly worse in general. 
In the 0\,dB case, the WER with iterative optimisation is more than halved compared to the baseline model.

In contrast to the other datasets, CHiME-4 does not offer a variable SNR. In our experiments, we report on two variants of the cold cascade. In the first case, the SE model is trained cross-corpus with the Librispeech dataset, in the second case, it is trained on the CHiME-4 training set. Note that the training of the cold cascade approaches only implies a training of the AE component, while the evaluation is based on the ASR models supplied by the challenge. Similarly, for the MTL and iterative training approach, we train an AE and ASR system on CHiME-4 as described in \Cref{sec:method}, but evaluate the SE system in combination with the provided ASR systems GMM-HMM and DNN-HMM.

\begin{table*}[ht!]
  \begin{center}
  \caption{Speech emotion recognition testing results, Unweighted Average Recall (UAR)[$\%$], using DEMoS and the AudioSet corpus. DA stands for the method using only data augmentation. MTL represents the proposed multi-task learning solution.}
  \renewcommand{\arraystretch}{1.2}
  \begin{tabular}{l c c c c c c c c}
    \toprule
    Methods & Inf & $25\si{\deci\bel}$ & $20\si{\deci\bel}$ & $15\si{\deci\bel}$ & $10\si{\deci\bel}$ & $5\si{\deci\bel}$ & $0\si{\deci\bel}$ & average\\
    \midrule
    original SER & $81.32$ & $81.18$ & $79.95$ & $78.98$ & $73.70$ & $59.82$ & $40.32$ & $68.99$\\
    DA & - & $79.53$ & $79.46$ & $79.05$ & $78.30$ & $75.69$ & $68.06$ & $76.68$\\
    Cold Cascade & - & $80.45$ & $79.59$ & $79.45$ & $77.86$ & $69.93$ & $54.34$ & $73.60$\\
    Cold Cascade + DA & $-$ & $77.59$ & $77.27$ & $77.07$ & $77.54$ & $74.52$ & $69.02$ & $75.49$\\
    \hline
    \textbf{MTL} & - & $81.30$ & $80.67$ & $80.31$ & $79.93$ & $77.29$ & $75.44$ & $79.16$\\
    \textbf{iterative optimisation} & - & $\textbf{81.31}$ & $\textbf{80.76}$ & $\textbf{80.35}$ & $\textbf{79.95}$ & $\textbf{78.09}$ & $\textbf{76.91}$ & $\textbf{79.56}$\\
    \bottomrule
  \end{tabular}
  \label{Results_Table_SER}
  \end{center}
  \vspace{-0.7cm}
\end{table*}

\Cref{tab:CHiME} shows the results for simulated (simu) and real test set with the two provided ML approaches GMM-HMM and DNN-HMM, as introduced in \Cref{sec:method}. 
The iterative optimisation achieves the lowest WER in all cases with MTL being the follow-up. The cross-corpus approach cold cascade 1 consistently performs worse than cold cascade 2, however, still outperforming the baseline. Overall, the WERs are comparable to the 25\,dB case of the Librispeech experiments with the best result being the iterative optimisation with a DNN-HMM on the real test data at a WER of $8.12\,\%$

\subsection{Downstream Task III: Speech Emotion Recognition}
Our method is evaluated for the task of \ac{SER} on a dataset of elicited mood in Italian speech, DEMoS~\citep{DEMOS}. 
DEMoS contains 9\,365 emotional and 322 neutral samples recorded from 68 native speakers (23 females and 45 males; mean age 23.7\,years, standard deviation 4.3\,years). 
Six emotions -- anger, sadness, happiness, fear, surprise, and guilt -- are elicited by listening to music, watching pictures or movies, pronouncing or reading emotional sentences, and recalling personal memories. 
Original recordings are captured at 44.1\,kHz with 16-bit depth.

In our experiments, we use all of the emotional samples in DEMoS, while keeping the partitioning of the data for training, development, and testing identical to \cite{9054087}, ensuring a speaker-independent split and accounting for gender and class balance. 
To be consistent with the other target audio applications, the audio samples from DEMoS are down-sampled to 16\,kHz, which yields no evident information loss according to \cite{9054087}.
As for the previous experiments, we simulate the background noise along with the speech utterances by adding environmental recordings from AudioSet\todo{+++}. 

The \ac{CNN} model predicts the emotion label given a single utterance as input.
To account for the imbalanced class distribution in the test set, we use \ac{UAR}, the unweighted average of the class-specific recalls, to evaluate the trained models throughout the experiments.

\Cref{Results_Table_SER} shows the results of our experiments.
On the clean test set, our model achieved a \ac{UAR} of $81.32\,\%$.
It is relatively robust to additive noise up to \ac{SNR} values of $15$\,\db; further increasing noise intensity over that leads to an average \ac{UAR} of $68.99\,\%$, far below the performance on the clean test set.
As for the baseline methods, data augmentation increases robustness, especially in low \ac{SNR} levels under $10$\,\db, but using a \ac{SE} frontend hardly improves, and even hampers, performance.

Both of our joint optimisation approaches indicate better SER performance, resulting in an average UAR of $79.16\,\%$ and $79.56\,\%$, surpassing the best performing baseline model (DA) which produces a UAR of $76.68\,\%$. 
This improvement partially arises from mitigating the language gap introduced by the cold cascade models, as the AE model is trained on data in English and the SER model in Italian, whilst additionally strengthening the integration of the two models.    

\begin{table*}[ht!]
  \begin{center}
  \caption{Acoustic scene classification testing results, (Acc)uracy[$\%$], using the DCASE2021 and Librispeech corpus. DA stands for the method using only data augmentation. MTL represents the proposed multi-task learning solution.}
  \setlength{\tabcolsep}{4pt}
   \renewcommand{\arraystretch}{1.2}
  \begin{tabular}{l c c c c c c c c c c}
    \toprule
    Methods & Inf & $-25\si{\deci\bel}$ & $-20\si{\deci\bel}$ & $-15\si{\deci\bel}$ & $-10\si{\deci\bel}$ & $-5\si{\deci\bel}$ & $0\si{\deci\bel}$ & $5\si{\deci\bel}$ & $10\si{\deci\bel}$ & average\\
    \midrule
    original ASC & $77.81$ & $75.45$ & $72.92$ & $69.05$ & $65.02$ & $60.14$ & $51.19$ & $39.12$ & $26.91$ & $57.48$\\
    DA & - & $70.51$ & $71.44$ & $71.50$ & $70.84$ & $69.08$ & $68.86$ & $63.73$ & $59.31$ &$68.16$\\
    Cold Cascade & - & $73.83$ & $73.11$ & $71.14$ & $67.60$ & $62.58$ & $56.87$ & $50.53$ & $44.53$ & $62.53$\\
    Cold Cascade + DA & $-$ & $72.92$ & $72.84$ & $72.65$ & $71.71$ & $69.27$ & $63.59$ & $60.16$ & $59.23$ &$67.80$\\
    \hline
    \textbf{MTL} & - & $\textbf{74.31}$ & $\textbf{73.97}$ & $73.01$ & $72.48$ & $71.52$ & $70.10$ & $65.79$ & $61.34$ & $70.32$\\
    \textbf{iterative optimisation} & - & $74.26$ & $73.50$ & $\textbf{73.12}$ & $\textbf{72.71}$ & $\textbf{72.09}$ & $\textbf{71.43}$ & $\textbf{66.81}$ & $\textbf{63.19}$ & $\textbf{70.89}$\\
    \bottomrule
  \end{tabular}
  \label{Results_Table_ASC}
  \end{center}
\end{table*}

\subsection{Downstream Task IV: Acoustic Scene Classification}
To test our approach on the final task, acoustic scene classification, we use the \ac{DCASE} 2021 Challenge dataset~\citep{Wang2021_ICASSP}.
To create the noisy scene audio, speech samples from LibriSpeech are added to the soundscape recordings from the \ac{DCASE} 2021 challenge. 
The range of \acp{SNR} is enlarged to -25, -20, -15, -10, -5, 0, 5, and 10\,\db, covering a wide range of real-life conditions.
We note that in computing \ac{SNR}, we still consider \emph{speech} as the `signal' and \emph{scene} as the `noise', to be consistent with other literature.
So a lower \ac{SNR} means that the scene is dominating the soundscape, while a higher \ac{SNR} means that the speech interference is greater.
Since the test data is balanced, we can use the standard classification accuracy as the evaluation metric, similar to the DCASE challenge.

Results are shown in \cref{Results_Table_ASC}.
Training and testing our \ac{ASC} model on the original recordings leads to a classification accuracy of $77.81\%$.
Increasing the interference caused by speech leads to severe degradation of performance with an average accuracy of $57.48\%$ over all SNR values, as was the case for all other tasks when increasing noise levels.
This illustrates why the \ac{ASC} task would also benefit from a denoising component.

Overall, our joint optimisation approaches yield higher average accuracies over all \acp{SNR}, with $70.89\,\%$ and $70.32\,\%$ over the best-performing baseline of $68.16\,\%$, respectively.
Our iterative optimisation is particularly suited to the high \ac{SNR} conditions, where speech dominates the soundscape, and its accuracy $63.19\,\%$ clealry surpasses that of our MTL method ($61.34\,\%$).
In most other cases, our optimisation approaches are near equivalent, while consistently outperforming all baselines.
We note that in this case, the baseline deteriorates to near chance-level performance, with $26.91\,\%$.

\section{Discussion}
\label{sec:Discussion}
Collectively, our results on four different application domains show that the proposed methods consistently outperform comparable baselines.
In particular, joint training, either in the form of MTL or in the form of iterative optimisation, yields better results than methods relying on data augmentation or cascade enhancement, which are standard baselines in the field.
This demonstrates that adapting an enhancement model to the downstream task can bring substantial improvements, which underlines the need for specialised \ac{AE} systems that are able to differentiate between the task-specific relevancy of audio signals and noise sources.

\begin{figure*}[ht!]
  \centering
    \includegraphics[scale=0.4]{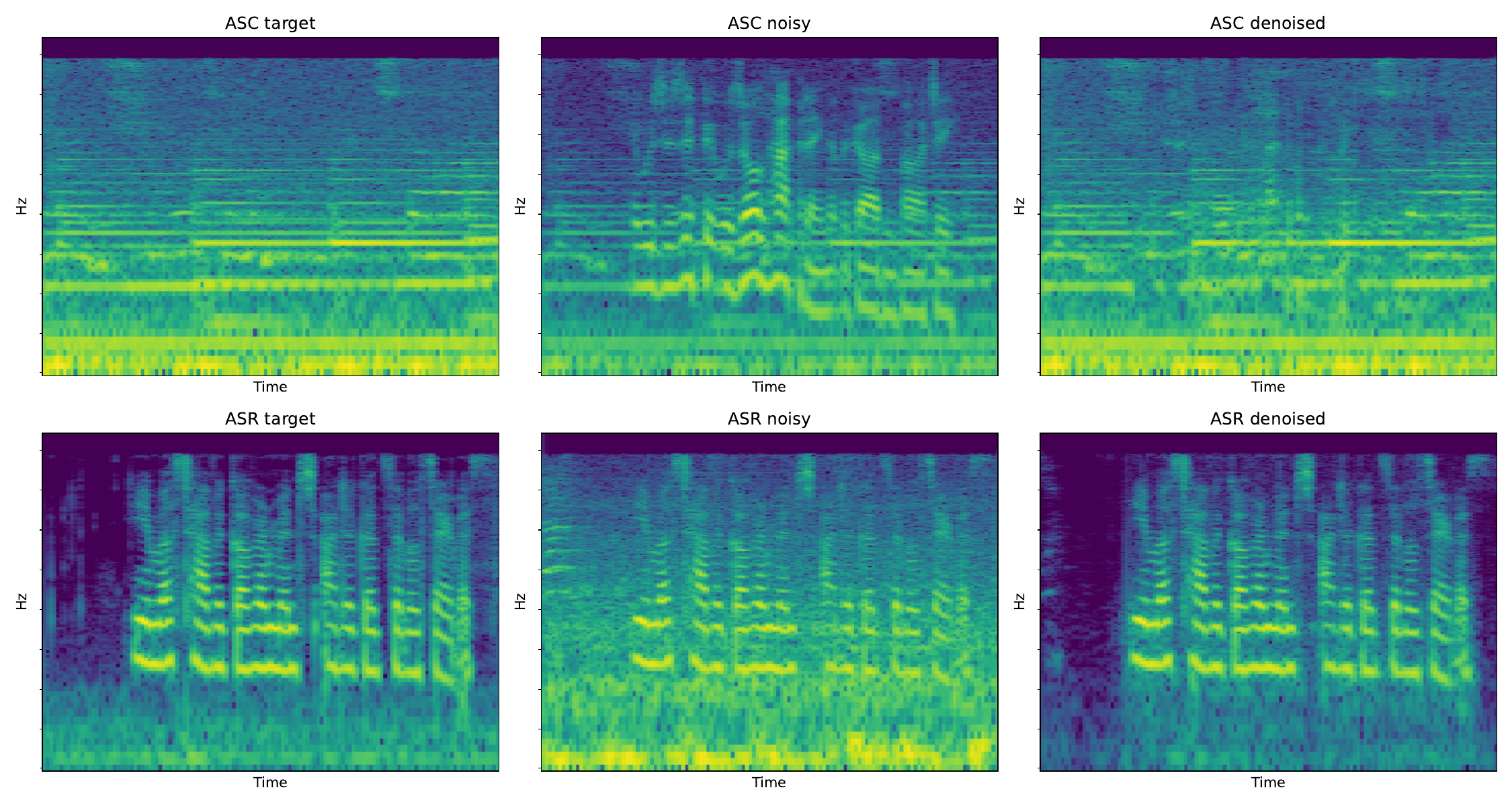}
    ~
  \caption{Spectrograms for visualisation of  audio enhancement samples. The left column displays the clean target samples, the middle column contains the artificially added noisy sample and the last column represents the reconstructed (denoised) audio. In the first row (audio scene classification) the clean sample is a sample of music and the considered noise is a speech sample, while in the second row (automatic speech recognition), the clean sample is a speech sample and the noise originates from a construction site.}\label{fig:spectrograms}
\end{figure*}

This aspect becomes even more apparent when comparing the audio enhancement output of the two tasks ASR and ASC depicted in Fig. \ref{fig:spectrograms}. For the ASC task (upper row) the clean audio can be described as background music with relatively stable frequency patterns over time. To construct the noisy sample, this is overlayed with a speech sample, which visually interrupts the smoothness of the background music. The AE module trained for ASC is thus capable of removing the disruptive speech signal and reconstructing the original target with only a few artefacts. In the bottom row, the clean sample is a speech sample, without any audible background noise. The spectrogram is thus characterised by an irregular frequency distribution --due to pauses and a change of frequencies between phonemes-- over a dark (quiet) background. After adding noise from a construction site, the frequency patterns of the speech samples stand out to a lesser degree from the stable background noise. The AE component trained on the ASR however is able to reconstruct the distinctiveness of the speech sample by suppressing the noise frequencies. 

Furthermore, our iterative optimisation is also consistently superior to MTL; it is only outperformed in very few of the high SNR conditions for speech command recognition and acoustic scene classification.
In most cases, it is able to recover a substantial percentage of the performance loss incurred by noise, even at 0\,\db: $\backsim69\,\%$ for SCR, $\backsim60\,\%$ for ASR, $\backsim\,89\%$ for SER, $\backsim\,71\%$ for ASC.
Since the audio applications we selected to test our method are associated with a broad range of real-life audio environments, we are optimistic that the system can be used in a wide range of other applications.

However, some limitations are attached to this study, as some baseline models have a performance lower than recent state-of-the-art methods when considering only training and testing on clean data. For instance, an ASR model can take advantage of self-supervised learning~\citep{zhu2022joint}, which allows them to scale up the amount of data and reap the benefits that this entails.
However, an SSL framework is agnostic to the task; thus, it cannot adjust the importance of individual samples based on downstream performance, which was found highly beneficial in our work.
This leads us to conclude that using joint SSL and enhancement pre-training on larger amounts of data, followed by fine-tuning with our iterative optimisation on the target downstream task is a promising avenue of future research.

Furthermore, beyond preliminary experiments, we have chosen to focus on only one type of denoising architecture (U-Net), which raises some concerns as to how well our approach would generalise to other models.
Nevertheless, as the proposed methods are agnostic to the underlying architectures, we expect them to show similar improvements when combined with other state-of-the-art models.

\section{Conclusion}
\label{sec:Conclusions}
In this work, we focused on single-channel audio enhancement adapted to specific computer audition downstream tasks under low signal-to-noise-ratio (SNR) conditions.
In particular, we considered the downstream tasks of speech command recognition (SCR), automatic speech recognition (ASR), speech emotion recognition (SER), and acoustic scene classification (ASC); in the first three, speech was the signal of interest and background soundscapes had to be removed; in the last case, this was reversed as now speech was the interfering source.
Instead of following a separate training paradigm for audio enhancement and downstream task models, we proposed the iterative optimisation method that increases the interplay between the two models in training.   
The testing results indicate considerable improvements measured by the respective evaluation metrics for each task, especially for low SNRs.
Our work shows that tailoring an audio enhancement front-end to the particular downstream task that requires denoising yields substantial improvements over generic enhancement models that have been trained on out-of-domain data.
Inspired by the improvements, more efforts should be put into further strengthening the coupling of the two models, for instance by exploring different weights for the sample importance, as well as utilising recent advances in self-supervised learning which have been shown to improve audio enhancement.

\section*{Acknowledgements}
This research has been partly supported by the Affective Computing
\& HCI Innovation Research Lab between Huawei Technologies and
the University of Augsburg. It was also partially funded from the EU H2020 project No.\ 101135556 (INDUX-R)

\bibliographystyle{plainnat}
\bibliography{2934-source}

\vspace{5mm}
\noindent\parbox{8.3cm}{\parpic{\includegraphics[width=9em]{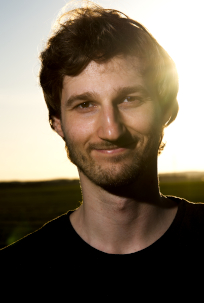}}{\small\quad {\bf Manuel Milling} received his Bachelor of Science in Physics and in Computer Science from the University of Augsburg in 2014 and 2015, respectively and his Master of Science in Physics from the same university in 2018. He is currently a PhD candidate in Computer Science at the chair of Health Informatics, Technical University of Munich. His research interests include machine learning with a particular focus on the core understandings of and applications of deep learning methodologies.
} }\\[1mm]

\noindent\parbox{8.3cm}{\parpic{\includegraphics[width=9em]{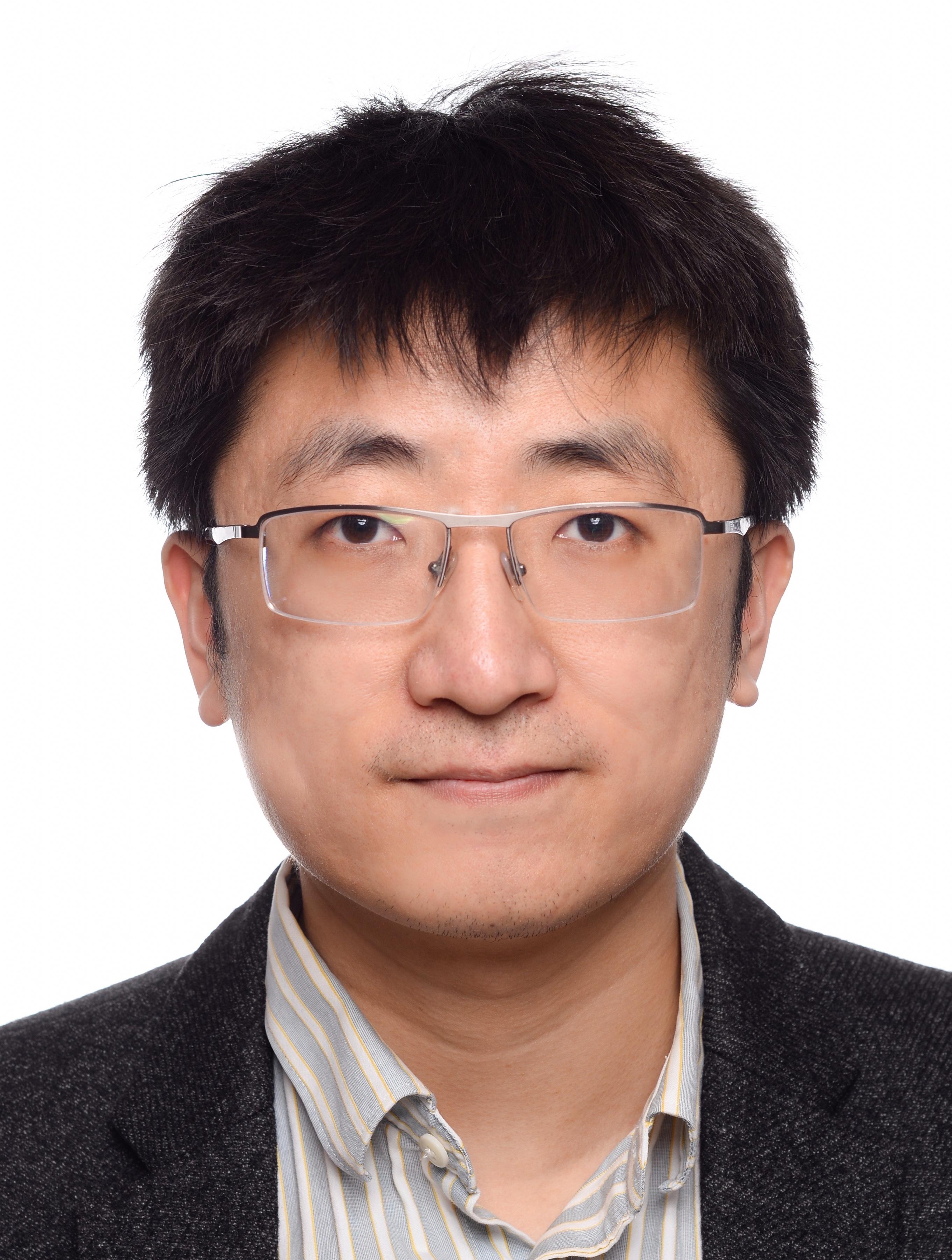}}{\small\quad {\bf Shuo Liu}  received his Bachelor degree from the
Nanjing University of Posts and Telecommnunica-
tions, China, in 2012, and his M.Sc. from the Tech-
nical University of Darmstadt, Germany, in 2017.
He worked as a researcher in the Sivantos group
for hearing aids solutions. He is currently a Ph.D.
candidate at the Chair of Embedded Intelligence for
Health Care and Wellbeing, University of Augsburg,
Germany. His research focuses are deep learning for
audio processing, mobile computing, digital health,
and affective computing.}\\[1mm]}

\noindent\parbox{8.3cm}{\parpic{\includegraphics[width=9em]{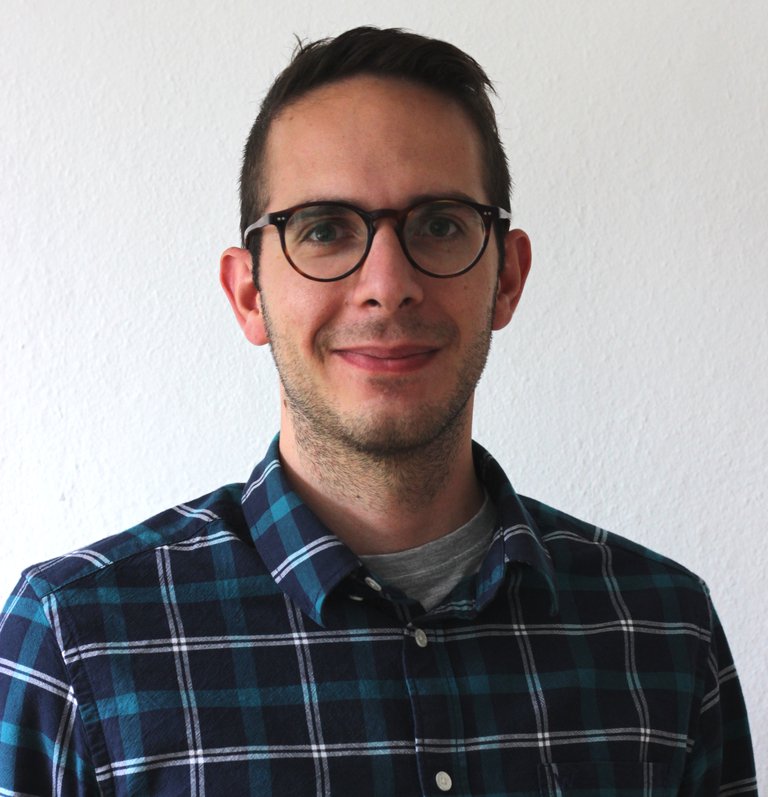}}{\small\quad {\bf Andreas Triantafyllopoulos} received the diploma in ECE from the Univer-
sity of Patras, Greece, in 2017. He is working toward
the doctoral degree with the Chair of Health Informatics, Technical University of Munich. His current focus is on deep learning methods for auditory intelligence and affective com-
puting.}\\[1mm]}

\noindent\parbox{8.3cm}{\parpic{\includegraphics[width=9em]{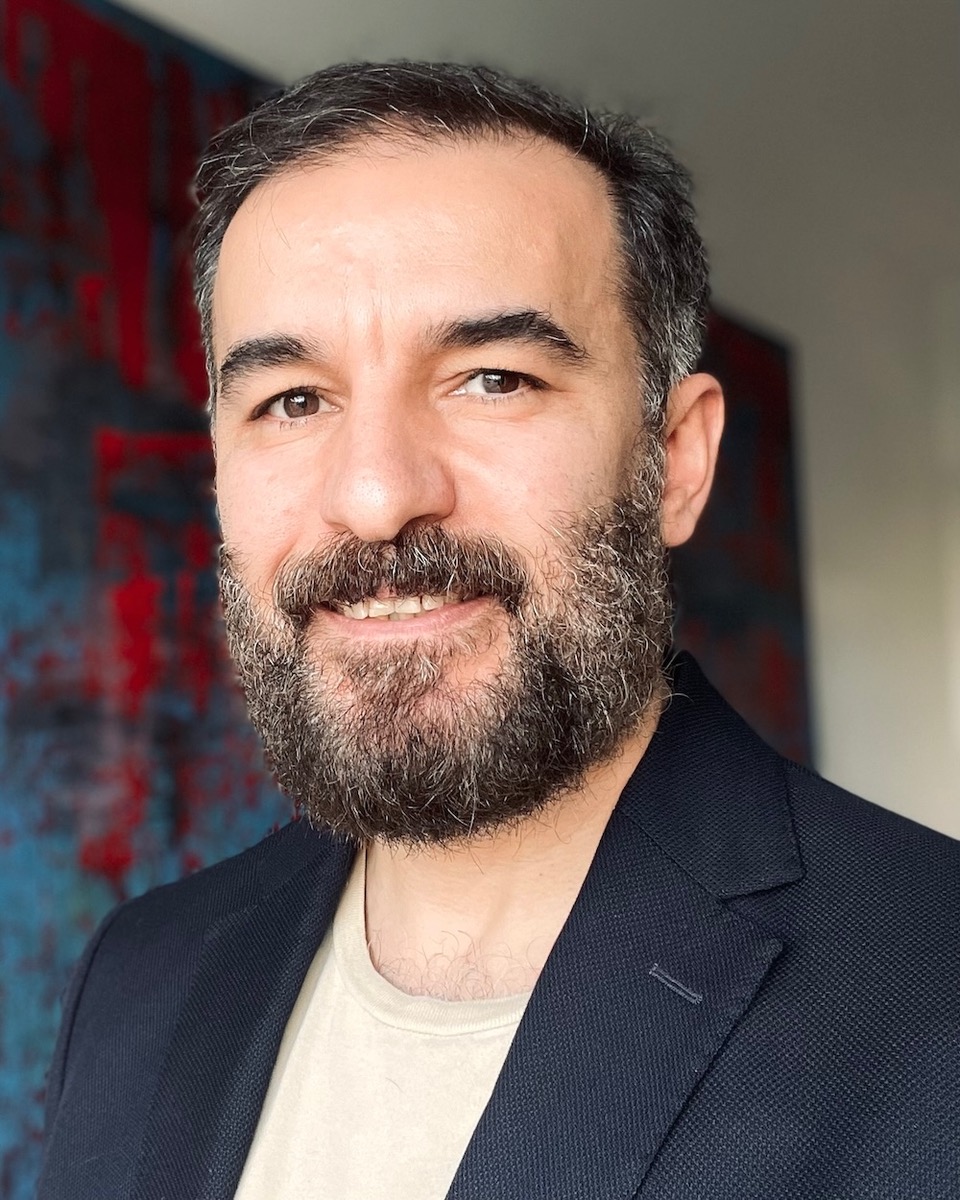}}{\small\quad {\bf Ilhan Aslan}  received the Diploma in 2004 from Saarland University in Germany and the Doctoral degree in 2014 at the Center for HCI from Paris-Lodron University Salzburg in Austria. He was an akad. Rat (assistant professor) at University of Augsburg from 2016 onward before joining Huawei Technologies in 2020 as an HCI Expert where he leads an HCI  research team and managed the Affective Computing \& HCI Innovation Research Lab. His research focus is at the intersection of HCI, IxD, and Affective Computing, exploring the future of human-centred multimedia and multimodal interaction.}\\[1mm]}

\noindent\parbox{8.3cm}{\parpic{\includegraphics[width=9em]{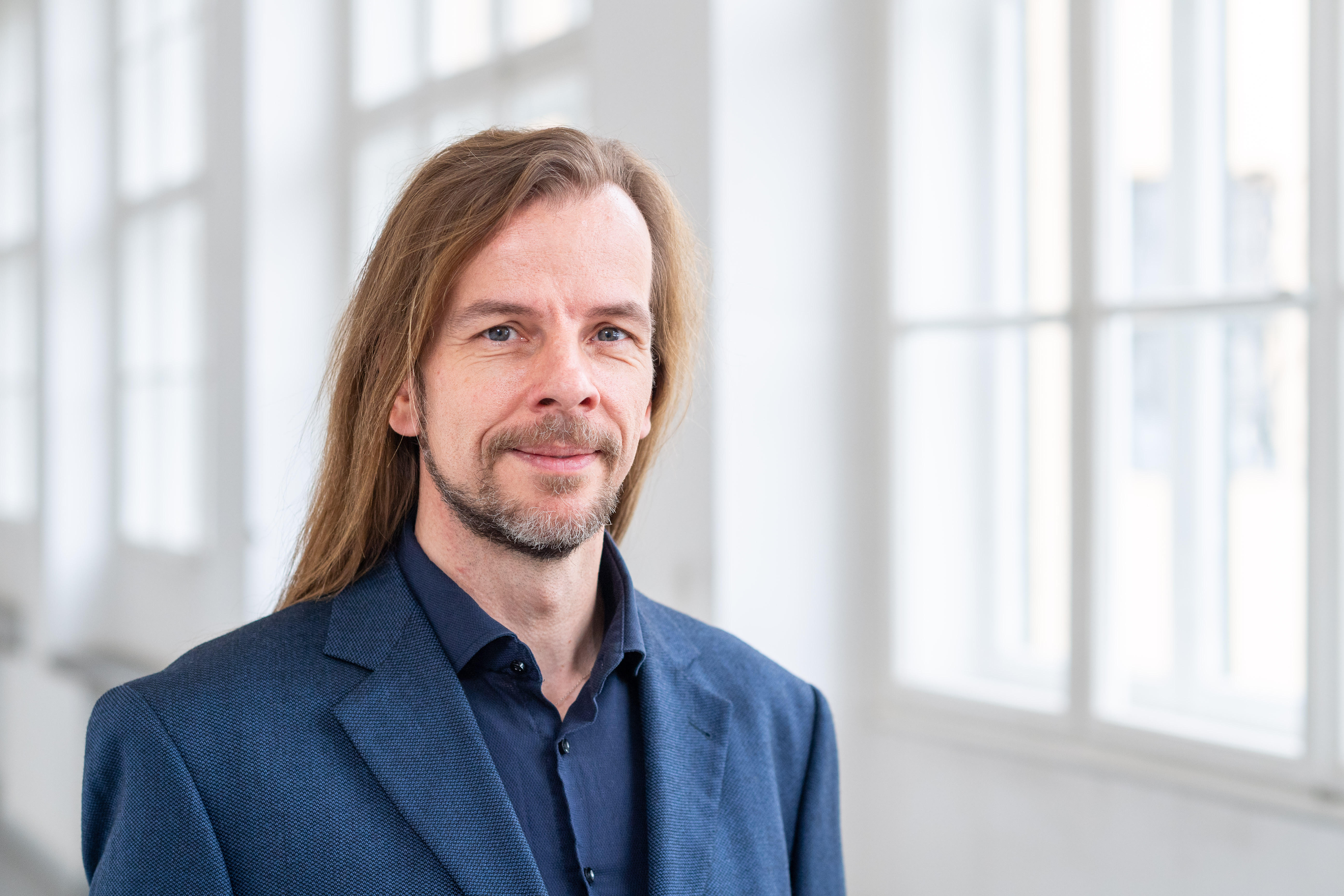}}{\small\quad {\bf Bj\"orn W\ Schuller} 
received his diploma in 1999,
his doctoral degree in 2006, and his habilitation and
was entitled Adjunct Teaching Professor in 2012 all
in electrical engineering and information technology
from TUM in Munich/Germany. He is Full Professor
of Artificial Intelligence and the Head of GLAM at
Imperial College London/UK, Chair of CHI -- the Chair
for Health Informatics, MRI, Technical University
of Munich, Munich, Germany, amongst other Professorships and Affiliations.
He is a Fellow of the IEEE and Golden Core Awardee of the IEEE Computer Society, Fellow of the ACM, Fellow and President-Emeritus of the AAAC, Fellow of the BCS, Fellow of the ELLIS, Fellow of the ISCA, and Elected Full Member Sigma Xi. He (co-)authored 1,400+ publications (60,000+ citations, h-index=110).}\\[1mm]}

\label{last-page}
\end{multicols}
\label{last-page}
\end{document}